# Charge Transport Properties of Molecular Junctions built from Dithiol Polyenes

Alexander Schnurpfeil[a,b]* and Martin Albrecht[b]

[a]*Institute for Theoretical Chemistry - University of Cologne - 50939 Köln/Germany,*

[b]*Theoretical Chemistry FB08 - University Siegen - 57068 Siegen/Germany*

## Abstract

We present a study of the charge transmission behavior of a series of dithiol polyenes in the context of molecular junctions. Using the Landauer theory and zero voltage approximation the Green's functions of the inserted molecules are calculated from a fully *ab initio* wave function based procedure. Various possibilities in approximating the correlation space are explored and quantitatively evaluated. Our results show that the transmission behavior of a molecular junction is not a monotonic function of the length of the employed molecule. Moreover we introduce the analytic solution of a suitable model system to countercheck the *ab initio* results and find a remarkable degree of correspondence.

* Corresponding author: e-mail: a.schnurpfeil@uni-koeln.de, Phone: +49 221 470 6885

## I. INTRODUCTION

Recent years have seen a steep rise in the broad field of nano engineering with molecular junctions being a significant part of it[1]. This has been brought about by tremendous advances in engineering techniques, which lead to both unraveled reduction in size and variety[2]. The particular interest in molecular junctions is to ultimately design switches or 'transistors' on a nano scale which might be triggered by various phenomena. The prototype molecular junction consists of two gold electrodes produced in break–junction experiments, which capture an organic molecule in between that can bind covalently to the electrodes via sulfide bridges[3].

The theory of charge transport through molecular junctions is, however, notoriously difficult. Most theoretical descriptions try to illuminate partial aspects like the role of the molecular electronic structure[4–6] or the influence of various structural conformations[7–9]. It is most prominently the set of methods based on the local density approximation (LDA) to density functional theory (DFT) as a starting point, which provides numerically affordable applications to the molecular junction problem[4–10] (cf. Ref.[4–8,10] for applications to carbon wires and benzene–based systems). Further approximations are commonly built on top of LDA, like the tight binding approach (TB) or parametrized minimal basis sets[11]. Another set of approaches renounces completely attempts of *ab initio* calculations and resorts to empirical models[12–15]. Most recently the DFT based augmented plane–wave method was applied to monowires by Mokrousov *et al.*[16]. Examples for further approximations are given by Guitierrez *et al.* in an application to an all–carbon–system with capped nanotubes as electrodes[11], Fagas *et al.* analyzing the off–resonant electron transport in oligomers[17] and Cuniberti *et al.* illuminating the role of the contacts[18,19]. An application predicting the actual current-voltage behavior of two aromatic molecules was demonstrated by Heurich *et al.*[20]. A more advanced scheme developed by Xue, Ratner and Datta sticks with these approximations, but develops a non–equilibrium formalism[13,15,21,22]. Earlier attempts were presented by Wang, Guo and Taylor.[23–25]. Principle ideas go back to Caroli *et al.* who originally focused on non–interacting systems[26]. While conceptually somewhat different, for the case of dc currents the ansatz of Cini is known to yield the same results[27]. The steady–state procedures have then been formulated as DFT schemes along the lines proposed by Lang[28]. However, possibly due to inherent shortcomings of the LDA approach, some results are far off experimental data. Calculations of Di Ventra *et al.* turned out to be off by two orders of magnitude[6,29], while different functionals are found to lead to large fluctuations of one order of magnitude[29,30]. The reason behind this is assumed to be the fact that transmission functions obtained from static DFT approaches have resonances at the Kohn–Sham eigenenergies, which frequently do not coincide with the physical excitation energies. In contrast, a most recent wave function based *ab initio* approach based on scattering theory came into the range of experiments for the same kind of organic systems[31].

Wave function based methods, on the other hand, are straightforwardly applicable to both ground state and excited state calculations alike. Subsequently, quantities like the self energy or the transport coefficient T can be obtained in a reliable manner. Furthermore they are amenable to systematic improvement on the numerical accuracy. However, the numerical demand increases tremendously with the system size. A demonstration of a wave function based calculation of the transport current through a molecular junction was recently given by the aforementioned work of Delaney and Greer[31]. Their original approach was directly derived from scattering theory, thus avoiding the calculation of any Green's function. The general bottle–neck of steep increase of numerical effort with system size, however, affects all wave function based methods alike. It is precisely this obstacle which can be overcome by a formulation of electron correlations in local orbitals and a hierarchy of correlation contributions called the incremental scheme. The usefulness of local orbitals has recently been confirmed in the frame of an LDA tight binding approach to transport properties of nanowires[32–34] as well as carbon nanotubes[35].

One of the authors has developed a wave function based *ab initio* method to obtain the Green's function of semiconductors[36–39]. The key enabling such calculations for solids was an approach based on local orbitals and a real space formulation of the self energy. In a recent development similar ideas were shown to bring about significant progress for the case of molecules. Applications have been performed for dithiolbenzene and its meta version[40,41]. Furthermore a first attempt to extend this approach to the non–equilibrium case has been made most recently[42,43], albeit in a minimal basis set.

In this paper we stick to the Landauer approximation applied earlier to dithiolbenzene[40,41]. Our procedure is applied to a series of dithiol polyenes with increasing chain length. The transmission coefficient T is obtained and discussed as a function of the chain length. Various ways of setting up the Green's function and the transmission coefficient are considered. The results are then compared to a model chain.

In Sec. II we briefly summarize the theory underlying our formalism. The numerical results obtained for the dithiol polyenes are presented in Sec. III and are confronted with the findings of a model system for which analytic formulas are computed in Sec. IV. Our conclusions are given in Sec. V.

## II. THEORY

### A. The Green's function

Our correlation scheme is formulated in terms of local occupied and local – or alternatively canonical – virtual HF orbitals. In the basis of these orbitals the one–particle Green's function is set up. As an example the case of virtual molecular states is discussed. The model space P describing the HF level then comprises of the $(N+1)$–particle HF determinants $|n\rangle$, while the correlation space Q contains single (and, in principle, double) excitations $|\beta\rangle$ on top of $|n\rangle$:

$$|n\rangle = c_n^\dagger |\Phi_{\text{HF}}\rangle, \qquad |\beta\rangle = c_r^\dagger c_a |n\rangle, \quad c_r^\dagger c_s^\dagger c_a c_b |n\rangle \tag{1}$$

$$P = \sum_n |n\rangle\langle n|, \qquad Q = \sum_\beta |\beta\rangle\langle \beta|. \tag{2}$$

In this local description indices provide an orbital index $n$ which is normally taken to indicate a local HF orbital and includes the spin index. We adopt the convention that indices $a, b, \ldots$ refer to occupied HF orbitals, $r, s, \ldots$ denote virtual orbitals and $m, n, \ldots$ can be either occupied or unoccupied orbitals. The idea of local orbitals given above translates into a restriction of the area the orbital can be chosen from to one or more contiguous spatial parts of the molecule. It is important to note that by enlarging the size of the spatial area thus covered this approximation can be checked in a systematic way for convergence. This leads to the incremental scheme introduced in Sec. II C.

Pertaining to the above notation the Green's function matrix $\mathbf{G}$:

$$G_{\text{nm}}(t) = -i\langle T[c_{\text{n}}(0) c_{\text{m}}^\dagger(t)]\rangle, \tag{3}$$

where T is the time–ordering operator and the brackets denote the average over the exact ground state, can be obtained from Dyson's equation as:

$$G_{nm}(\omega) = [\omega \mathbf{1} - \mathbf{F} - \mathbf{\Sigma}(\omega)]_{\text{nm}}^{-1}. \tag{4}$$

Here the self energy $\mathbf{\Sigma}(\omega)$ which contains the correlation effects, has been introduced and $\mathbf{1}$ is the identity matrix. The correlated eigenenergies are given by the poles of the Green's function which are numerically iteratively retrieved as the zeros of the denominator in Eq. (4). The density of states and satellites can also be obtained from **G**. To construct the self energy the resolvent

$$\left[\omega - H^{\text{R}} + i\delta\right]^{-1}_{\beta;\beta'} \tag{5}$$

is needed. It can be gained from diagonalization of the Hamiltonian

$$[H^{\text{R}}]_{\beta,\beta'} = \langle\beta|H - E_0|\beta'\rangle, \tag{6}$$

where the states $|\beta\rangle, |\beta'\rangle$ are those of the correlation space Q as in Eq. (1).

A decisive boost in numerical efficiency was brought about by combining the method described with the aforementioned incremental scheme formulated in local HF orbitals[36,37,44,45].

At this point we briefly introduce the notion of a molecular junction. Fig. 1 visualizes the concept. A molecule (such as a dithiol polyene) is fixed between two gold electrodes. The sulfur of the thiol groups has a high affinity to gold and will build a covalent type of bonding to the electrodes. The molecule can thus be regarded as a conductor (junction) conducting the current from one electrode to the other.

### B. The Landauer Theory

From the Green's function, the transport coefficient $T$ can be straightforwardly obtained in the frame of the Landauer formalism[46]. This theory constitutes an approximation, assuming zero voltage across the junction, which is frequently adopted and finds its justification in the zero-current theorem[14].

In this context $T$ is given by

$$T = \text{Tr}\left\{\mathbf{\Gamma}_{\text{L}}\,\mathbf{G}\,\mathbf{\Gamma}_{\text{R}}\,\mathbf{G}^{\dagger}\right\}, \tag{7}$$

$$\mathbf{\Gamma}_{\alpha} = i[\mathbf{\Sigma}_{\alpha} - \mathbf{\Sigma}_{\alpha}^{\dagger}], \tag{8}$$

$$\mathbf{\Sigma}_{\alpha} = \mathbf{H}_{\text{M}\alpha}\mathbf{G}^{0}_{\alpha\alpha}\mathbf{H}_{\alpha\text{M}}, \tag{9}$$

$$\alpha = \text{L, R}, \tag{10}$$

where the indices L,R refer to the coupling to the left (L) and the right (R) electrode and can be obtained from the self energies of the respective coupling regions as shown in Eq. (8). The self energies in turn are obtained in a partitioning approach as depicted in Eq. (9). Here, the index M refers to the molecule. Eq. (9) requires in principle the exact knowledge of the isolated lead surface Green's function $\mathbf{G}^0_{\alpha\alpha}$ for both sides of the junction ($\alpha = \mathrm{L, R}$). In the wide band approximation, which has been adopted throughout this work, the coupling self energies provide an overall widening of the molecular energy levels, in particular at the sulfide bridges, due to the interaction with the energy continuum provided by the metal[5]. This approximation introduces a coupling between molecule and electrodes parameterized by a coupling constant $\delta$ which replaces the evaluation of Eq. (9).

The Green's function matrix $\mathbf{G}$ represents the entire system and is to be obtained from a partition approach leading to:

$$\mathbf{G}_{\mathrm{MM}} = \left[ {\mathbf{G}^0_{\mathrm{MM}}}^{-1} - \mathbf{\Sigma}_{\mathrm{L}} - \mathbf{\Sigma}_{\mathrm{R}} \right]^{-1}, \qquad (11)$$

where the superscript 0 refers to the Green's function of the bare molecule, obtained from the *ab initio* incremental scheme to be discussed below in Sec. II C. (The scheme is displayed for the key quantity T, but has been shown in earlier applications to also hold for the self energy $\Sigma$, hence for the Green's function itself, cf. Ref.[36,37,44]).

### C. The incremental scheme

As we demonstrated in earlier applications[40,41], the Green's function and hence the transmission coefficient T can be obtained in a step–wise manner by applying the so called incremental scheme.

As an illustration of the incremental scheme we choose trans–1,2–dithiolbutadiene which is depicted in Fig. 2. As subsets of the system some arbitrary spatial parts of the molecule, representing a suitable partitioning, are chosen. In the figure the thiol groups are summarized as regions I and IV, respectively, while regions II and III comprise of a $H - -C = C - -H$ group each. An incremental description of the transmission coefficient $T$ could start with a correlation calculation, in which only excitations inside one of the regions I–VI, *e. g.* region I, are allowed. This results in a contribution to the correlation correction to the self energy, and ultimately to the transmission coefficient T, which is labeled by the region it refers to, *e. g.* $\Delta T^{\mathrm{I}}$:

$$\Delta T^{\mathrm{I}} = T^{\mathrm{I}}. \qquad (12)$$

The sum of all one–region increments yields the so called one–region increment approximation (denoted $S$) to the transmission coefficient. In a next step the calculation is repeated with excitations correlating the charge carriers being allowed inside two regions *e. g.* region II and region III as shown in the middle of Fig. 2. The difference of this extended calculation $T^{\text{II,III}}$ with respect to the one–region increments $\Delta T^{\text{II}}$ and $\Delta T^{\text{III}}$ then isolates the effect of additional excitations involving the extended region consisting of region II and region III and forms a two–region increment $\Delta T^{\text{II,III}}$:

$$\Delta T^{\text{II,III}} = T^{\text{II,III}} - \Delta T^{\text{II}} - \Delta T^{\text{III}} \qquad (13)$$

In particular, this increment will be denoted as next–neighbor–two–region increment ($nD$). In general two–region increments are symbolized by $D$.

This procedure can be continued to more and more regions and the naming of this increments is analogous. Eq. (14) for example depicts a three–region increment, which will be denoted as $T$ throughout the remainder of this work, whereas $M$ represents results where the entire molecule is included in one increment. In the end the summation Eq. (15) of all increments is the final approximation to the sought transmission coefficient, *e. g.*:

$$\Delta T^{\text{I,II,IV}} = T^{\text{I,II,IV}} - \Delta T^{\text{I,II}} - \Delta T^{\text{I,IV}} - \qquad (14)$$
$$\Delta T^{\text{II,IV}} - \Delta T^{\text{I}} - \Delta T^{\text{II}} - \Delta T^{\text{IV}}$$

$$\boxed{\begin{aligned} T = \quad & \sum_{A=I}^{IV} \Delta T^{A} + \\ & \sum_{A>B=I}^{IV} \Delta T^{A,B} + \\ & \sum_{A>B>C=I}^{IV} \Delta T^{A,B,C} + \\ & \qquad \ldots \end{aligned}} \qquad (15)$$

From the experience gained with the incremental scheme in its application to the self energy, a rapid decrease of increments both with the distance between the regions involved and with their number included in the increment can be expected. This means that only a few increments might need to be calculated. It is crucial to emphasize that the cutoff thus introduced in the summation

Eq. (15) is well controlled, since the decrease of the incremental series can be explicitly monitored. The validity of this point is demonstrated in the discussion of the results. There we will also see that the convergence behavior of the transmission coefficient is even better than the convergence behavior of the correlation correction to excitation energies.

### III. RESULTS AND DISCUSSION

We start our discussion of the results with the obtained correlation corrected HOMO-LUMO gap of the dithiol polyene oligomers, which we refer to as chains henceforth. In order to shorten the notation, we introduce the abbreviation $\gamma = (LUMO - HOMO)_{HF} - (LUMO - HOMO)_{CORR}$, where $(LUMO - HOMO)_{HF}$ denotes the HOMO-LUMO gap on the HF level and $(LUMO - HOMO)_{CORR}$ donates the correlated gap. The geometrical structure of the dithiol polyene chains was optimized on the B3LYP–level of DFT with a 6-31G(d) basis set using the quantum chemistry program GAUSSIAN03[47]. The results of $\gamma$ are given in Tab. I. We started the series of oligomers with trans-1, 2-dithiolethylene and increased the chain length step by step with $H-C=C-H$-fragments. The number of such fragments will be denoted by $\eta$. In order to avoid a lengthy notation we abbreviate the molecules by their number of $H-C=C-H$-fragments. For instants 1, 2-dithiolethylene will be referred to as *molecule 1*. The first column denotes the molecule under consideration, the second column shows the increments which were used in the calculations in order to get the correlation energies. The calculations were performed in two different ways: The third column in Tab. I represents the results when the entire canonical virtual space – henceforth denoted as *space 1* – was included in the perturbation calculation. The fourth column contains the results when the virtual space was localized by the method of Pipek and Mezey[48] – henceforth denoted as *space 2* – using the quantum chemistry package MOLPRO[49]. In both cases – canonical virtual space and localized virtual space – the occupied orbitals were localized by the same method. In the case of *space 1* the whole virtual space is assigned to each increment while in the case of *space 2* only those localized virtual orbitals are included which lie in the particular region of the increment under consideration. For the calculation of the $\gamma$'s and T's a $cc-pVDZ$ basis set was used. The difference of the numerical effort which results in using *space 1* and *space 2* respectively in the calculations is remarkable. While the entire virtual space is included with each increment by using *space 1*, the virtual space is increased step by step with the use of more and more increments using *space 2* until the total virtual space is switched on if the entire molecule is considered as one increment. So it is obvious that the correlation contribution one obtains *e. g.* with a one–region increment is closer to a complete calculation when *space 1* instead *space 2* is applied because the correlation space is much larger. The influence of different correlation spaces can be realized by considering the amount of $\gamma$ when the entire molecule is included in the calculation. For *molecule 1* we obtain 1.839 eV when the virtual space is localized in contrast to 3.032 eV when using the canonical virtual space. This error consecutively becomes smaller and smaller as the chain becomes larger. The difference between these two values amounts to 1.193 eV for *molecule 1* and decreases monotonously to 0.566 eV for *molecule 5*. This is an obvious manifestation of the transition from an oligomer to a polymer: as the oligomer increases in length, including the full virtual space loses some of its advantage as more and more long distance excitations are accounted for which however do not contribute to the correlation corrections. A obvious manifestation of the oligomer–polymer transition is the decrease of the gap with increasing chain length.

We now focus on the application of the incremental scheme. When *space 1* is used the table suggests that the main part of $\gamma$ can be obtained by computing only the one–region increments $S$. The difference between $\gamma$ obtained by considering the entire *molecule 1* compared to $\gamma$ obtained by using the $S$-increments amounts to merely 0.132 eV. In the case of *molecule 2* this difference becomes 0.453 eV. For the remaining molecules the deviations stay in the same range, *e. g.* 0.424 eV for *molecule 5*. By contrast, when considering the results computed by using *space 2* it is not possible to obtain the main part of $\gamma$ by only computing the $S$-increments. For example, the difference in the values for *molecule 2* amounts to 1.288 eV, which is significant. It can thus be concluded that it is necessary to use *space 1* in order to get reliable results concerning $\gamma$ while applying a strict cut in the incremental scheme. So we continue our discussion about the incremental scheme by only considering the results obtained with *space 1*. As mentioned before, it is possible to obtain the main part of $\gamma$ if just the $S$-increments are computed. To go a step further one can include a different number of two–region increments $D$ to improve the results. The more increments are included the more accurate the result will be. However, a close look at the values in Tab. I allows to establish that it is sufficient to include only the $S$- and $D$-increments in order to get satisfying results. In some cases $\gamma$ will be a little bit overestimated. This is an effect which is inherent to the incremental scheme. By including more and more increments this effect will be compensated. It is an important feature of this approach that reliable results can be obtained although the incremental scheme will be truncated. This is a decisive factor which allows to avoid the otherwise formidable computational costs inherent to an *ab initio* treatment. The calculation of $\gamma$ serves as a check for the incremental scheme which we predominantly used for the calculation of the transmission coefficient. Similar trends in the converging behavior are observed for the transmission coefficient T. Fig. 3 shows the devolution of the transmission coefficient with increasing coupling constant $\delta$ for *molecule 4*. When no coupling is assumed the transmission coefficient is zero which is in accordance with the physical intuition. The larger the coupling constant, the larger is the transmission coefficient which has its maximum when a coupling constant of 13 eV is applied. After this point the transmission coefficient is decreasing and tends to zero when a coupling constant of 27 eV is assumed. This behavior can be explained as follows: With an increase in the coupling the density of states (DOS) will also leak into the HOMO-LUMO gap of the bare molecule. This is demonstrated in the case of a rather week coupling of 3 eV in Fig. 4, where the various energy levels of the bare molecule are still discernable, but are clearly broadened somewhat from their otherwise perfect $\delta$–peak shape due to the coupling constant. At the same time some density of states becomes available in the HOMO–LUMO gap. So the probability of an electron transmission from the valence state to the unoccupied state will also be increased which results in a higher transmission coefficient. Upon excessive increase of this broadening, density will then be transfered into the regions far away from the gap, so that the charge density will be depleted again in the energy region of the original HOMO–LUMO gap, and hence the transmission coefficient decreases as well. The most remarkable fact is the fast convergence of the transmission coefficient when the incremental scheme is used. Fig. 3 displays the results if different kinds of increments are used for the computation: The solid line represents the transmission coefficient T in dependence of the coupling constant $\delta$ for the case that the entire molecule has been correlated. If T is approximated by the one–region increments $S$, the error resulting from this restriction on the incremental scheme will lead to just a slight overestimation given in percent by the dashed line. On average the error is just 1% with a peak of 6%. Upon inclusion of the two–region increments $D$ this error is reduced basically to zero as can be seen by the dotted line in the figure. In conclusion this means that it is sufficient to only compute the one–region increments in order to get reliable results for the transmission coefficient. The convergence properties are even better compared to the convergence properties of $\gamma$. This finding allows to save considerable numerical cost. Thus we are lead to believe that in the future it is feasible to perform calculations on significantly larger systems.

In Fig. 5 the transmission coefficients with varying coupling constants are depicted from *molecule 1* to *molecule 5* for the sake of comparison. For this calculation *space 1* was used. As expected *molecule 1* has the highest transmission coefficient T compared to the other dithiol polyenes under consideration. The maximum for this molecule occurs at a coupling constant of 11 eV. The same coupling constant gives the biggest value of the transmission coefficient of *molecule 2* which has the second highest transmission coefficient T, as expected. However, as we continue beyond *molecule 3*, the transmission coefficient increases again, contrary to intuition, and the maximum value is shifted to a higher value of the coupling constant, specifically 18 eV. Intuition might suggest that the conduction properties of *molecule 3* are somewhat better compared to *molecule 4* and *molecule 5* but the reverse is found. The maxima of the transmission coefficients of *molecule 4* and *molecule 5* lie above the values of *molecule 3* in the same range of the coupling constant of *molecule 1* and *molecule 2*. In order to explain this anomalous behavior the following aspects are of importance. Firstly the size and hence the length of the molecule and plays a decisive role. The smaller the molecule between the electrodes the higher tunneling–like effects across the molecule might be. This helps to understand the relatively high conduction characteristics of *molecule 1* and *molecule 2* and the reduced conduction qualities of *molecule 3*. A second aspect is the conjugation of the $\pi$-system of these molecules. As the chain length increases, the $\pi$–system gets ever more delocalized, thus facilitating conduction.

Finally the situation in energy space is also of interest. With increasing size of the molecule the HOMO–LUMO gap decreases, thus again facilitating charge transport. To some extent this last point is a rephrasing of the second, as it is the delocalization of the $\pi$–system which leads to a gap reduction with the chain length.

In this respect it is instructive to look at *molecule 4* and *molecule 5*. Here we observe an increase of the transmission coefficient compared to *molecule 3*. This results presumably from the increasing conjugation effect of the $\pi$–systems. Since the highest value of the transmission coefficient of *molecule 5* is even higher than the highest value of the transmission coefficient of *molecule 4* one can assume that the conjugation effect overcompensates the decreasing conducting effects which follow from the increase of the system dimensions.

We proceed our discussion by considering the computed transmission coefficients of the dithiol polyenes using *space 1* and *space 2* respectively. The results are given in Fig. 6. A coupling constant of 3 eV is assumed. Concerning *space 1*, the upper curve in Fig. 6 serves as a vertical cut of the diagrams in Fig. 5 at $\delta = 3$ eV. The lower curve in Fig. 6 shows the transmission coefficients when *space 2* is used rather than *space 1* in the calculations. The values are smaller because the correlation space is too small in order to assess all significant contributions to the transmission as discussed above. This behavior suggests that inclusion of the full canonical virtual space in the calculation is advisable so as to get converged results. We take this as a strong indication for the importance of the conjugated $\pi$–system, which evidently only is accounted for if the full virtual space is included.

At present we were not able to go to longer chains due to the increase in numerical cost. To further illuminate the above findings, we therefore introduce a simple model which allows for analytic solutions for any chain–length in the following section.

## IV. MODEL CALCULATIONS

In order to countercheck the quantum chemical results of the previous section we have set up an exactly solvable model system. It contains n-2 identical units $U_i, i \in \{2, 3, \cdots, n-1\}$, each of which contributes a virtual orbital d with energy $\epsilon_d$. The units are thought of to be arranged in the way of a one–dimensional chain. In addition at the left and right end we add another such unit with indices 1 and n, respectively. These are, contrary to the rest, coupled to electrodes. As in the previous sections the coupling is taken into account in the frame of the wide band approximation, so that the energy levels of these two units are broadened by an amount $\delta$, which is the external coupling constant.

We then apply a tight–binding scheme so that each unit only interacts with its two neighbors with an interaction strength t.

### A. The bare chain model

First the problem of a bare chain without contacts to a reservoir is considered. The Green's function

$$\mathbf{G}(\omega) = \begin{pmatrix} g_{11} & g_{12} & \cdots & g_{1n} \\ g_{21} & g_{22} & \cdots & g_{2n} \\ \vdots & & & \\ g_{n1} & g_{n2} & \cdots & g_{nn} \end{pmatrix} \tag{16}$$

describing this system must then fulfill the equation:

$$\begin{pmatrix} g_{11} & g_{12} & \cdots & g_{1n} \\ g_{21} & g_{22} & \cdots & g_{2n} \\ \vdots & & & \\ g_{n1} & g_{n2} & \cdots & g_{nn} \end{pmatrix} (-t) \begin{pmatrix} \frac{\omega-\epsilon_d}{-t} & 1 & 0 & 0 & \cdots & 0 \\ 1 & \frac{\omega-\epsilon_d}{-t} & 1 & 0 & \cdots & 0 \\ 0 & 1 & \frac{\omega-\epsilon_d}{-t} & 1 & \cdots & 0 \\ \vdots & & & & & \\ 0 & 0 & 0 & \cdots & 1 & \frac{\omega-\epsilon_d}{-t} \end{pmatrix} = \begin{pmatrix} 1 & 0 & 0 & \cdots & 0 \\ 0 & 1 & 0 & \cdots & 0 \\ \vdots & & & & \\ 0 & 0 & 0 & \cdots & 1 \end{pmatrix}. \quad (17)$$

In the following we subsume the parameter t into the Green's function ($g_{ij} \to -t g_{ij}$) and use the abbreviation $\Omega := \frac{\omega-\epsilon_d}{-t}$. Writing out this matrix equation just for the elements of the first line of the Green's function then yields the system of equations:

$$g_{11}\Omega + g_{12} = 1 \quad (18)$$

$$g_{11} + g_{12}\Omega + g_{13} = 0 \quad (19)$$

$$g_{12} + g_{13}\Omega + g_{14} = 0 \quad (20)$$

$$\vdots$$

$$g_{1,n-1} + g_{1,n}\Omega = 0,$$

which can be rewritten as

$$g_{10} + g_{11}\Omega + g_{12} = 0 \quad (21)$$

$$g_{11} + g_{12}\Omega + g_{13} = 0 \quad (22)$$

$$g_{12} + g_{13}\Omega + g_{14} = 0 \quad (23)$$

$$\vdots$$

$$g_{1,n-1} + g_{1,n}\Omega + g_{1,n+1} = 0 \quad (24)$$

with the agreement

$$g_{10} := -1, \quad (25)$$

$$g_{1,n+1} := 0, \quad (26)$$

so that all equations have the form:

$$g_{1,j-1} + g_{1,j}\Omega + g_{1,j+1} = 0, \quad j \in \{1, 2, .., n\}. \quad (27)$$

This leads to the iteration scheme

$$g_{1,j+1} = -g_{1,j}\Omega - g_{1,j-1}. \quad (28)$$

In particular we have

$$g_{1,n+1} = -\Omega g_{1,n} - g_{1,n-1}. \tag{29}$$

Replacing $g_{1,n}$ with the recursion (28) results in

$$g_{1,n+1} = \left(\Omega^2 - 1\right) g_{1,n-1} + g_{1,n-2}. \tag{30}$$

Continued iteration yields

$$g_{1,n+1} = -\Omega \left(\Omega^2 - 2\right) g_{1,n-2} - \left(\Omega^2 - 1\right) g_{1,n-3} \tag{31}$$

$$= \left(\Omega^4 - 3\Omega^2 + 1\right) g_{1,n-3} + \Omega \left(\Omega^2 - 2\right) g_{1,n-4}. \tag{32}$$

In general we find

$$g_{1,n+1} = \kappa_j g_{1,n-j} - \kappa_{j-1} g_{1,n-j-1}, \tag{33}$$

where we have introduced the $\kappa$–polynomials

$$\kappa_j := (-1)^{j+1} \left[ \Omega^{j+1} - j\Omega^{j-1} + (j-2)\Omega^{j-3} - (j-4)\Omega^{j-5} \cdots \begin{cases} (-1)^{\frac{j}{2}} 2\Omega & j \text{ even} \\ & \text{for} \\ (-1)^{\frac{j+1}{2}} & j \text{ odd} \end{cases} \right]$$

$$= (-1)^{j+1} \left[ \Omega^{j+1} + \sum_{l=1}^{L} (-1)^l (j - 2l + 2)\Omega^{j-2l+1} \right], \tag{34}$$

$$j + 1 =: 2L + r, \quad r \in [0, 1]$$

$$\Leftrightarrow L = \frac{(j+1) - mod(j+1, 2)}{2}.$$

Here $j + 1$ has been expressed in terms of multiples $L$ of 2 plus a remainder $r$ by means of the conventional function $mod(m, n)$.

Choosing $j = n - 1$ in Eq. (33) yields

$$g_{1,n+1} = \kappa_{n-1} g_{1,1} - \kappa_{n-2} g_{1,0}, \tag{35}$$

and upon inserting Eqs. (25,26) this is solved as

$$g_{1,1} = -\frac{\kappa_{n-2}}{\kappa_{n-1}}. \tag{36}$$

Going backwards in the system of coupled equations (28) yields for the remaining elements of the first line of the Green's function matrix $g_{1,m}$:

$$g_{1,2} = -\Omega \frac{\kappa_{n-2}}{\kappa_{n-1}} + 1, \qquad (37)$$

$$g_{1,m} = \kappa_{m-2} \frac{-\kappa_{n-2}}{\kappa_{n-1}} + \kappa_{m-3}. \qquad (38)$$

The remaining elements of the Green's function can be obtained in an analogous fashion. For the second line the system of equations (21–24) takes a similar form:

$$g_{21}\Omega + g_{22} = 0 \qquad (39)$$

$$g_{21} + g_{22}\Omega + g_{23} = 1 \qquad (40)$$

$$g_{22} + g_{23}\Omega + g_{24} = 0 \qquad (41)$$

$$\vdots$$

$$g_{2,n-1} + g_{2,n}\Omega + g_{2,n+1} = 0. \qquad (42)$$

Again the agreement

$$g_{2,n+1} := 0 \qquad (43)$$

was used. To obtain the same structure as for the previous case $g_{1,m}$ the second line Eq. (40) is rewritten as:

$$\tilde{g}_{21} + g_{22}\Omega + g_{23} = 0, \qquad (44)$$

$$\tilde{g}_{21} := g_{21} - 1, \qquad (45)$$

which in turn allows to rewrite Eq. (39) as:

$$\tilde{g}_{20} + \tilde{g}_{21}\Omega + g_{22} = 0, \qquad (46)$$

$$\tilde{g}_{20} := \Omega = -\kappa_0. \qquad (47)$$

The system of equations (39–42) can now be solved in the same way as was done above for the system (21–24). The analysis can be continued in the same fashion for all lines $j$ with elements $g_{j,m}$ of the Green's function.

The solutions are found to be:

$$\tilde{g}_{j,m} = \kappa_{j-2} g_{1,m}, \tag{48}$$

$$g_{j,m} = \tilde{g}_{j,m}, \qquad m \geq j,$$

$$g_{j,m} = \tilde{g}_{j,m} + \kappa_{j-m-2}, \qquad m \leq j.$$

With the convenient definition

$$\kappa_k := 0, \qquad k \leq -2 \tag{49}$$

the general solution is finally given by:

$$g_{j,m} = \kappa_{j-2} \left[ \kappa_{m-2} \frac{-\kappa_{n-2}}{\kappa_{n-1}} + \kappa_{m-3} \right] + \kappa_{j-m-2}. \tag{50}$$

As an illustration the problem for $n = 3$ is considered. With the solution (50) and the explicit form of the $\kappa$–polynomials (34,49) the problem (17) takes the form:

$$\frac{1}{\Omega^2 - 2} \begin{pmatrix} \frac{\Omega^2-1}{\Omega} & -1 & \frac{1}{\Omega} \\ -1 & \Omega & -1 \\ \frac{1}{\Omega} & -1 & \frac{\Omega^2-1}{\Omega} \end{pmatrix} \begin{pmatrix} \Omega & 1 & 0 \\ 1 & \Omega & 1 \\ 0 & 1 & \Omega \end{pmatrix} = \begin{pmatrix} 1 & 0 & 0 \\ 0 & 1 & 0 \\ 0 & 0 & 1 \end{pmatrix} \checkmark, \tag{51}$$

which indeed works out to be correct.

### B. The connection to the electrodes

Once the presence of the electrodes is taken into account in the frame of the wide band approximation and the Landauer theory, a slight change to the previous equations has to be added. In fact the Hamiltonian underlying the system changes as:

$$\mathbf{H} := \begin{pmatrix} \Omega & 1 & 0 & \cdots & 0 \\ 1 & \Omega & 1 & \cdots & 0 \\ \vdots & & & & \\ \cdots & 0 & 0 & 1 & \Omega \end{pmatrix} \to \tilde{\mathbf{H}} := \begin{pmatrix} \tilde{\Omega} & 1 & 0 & \cdots & 0 \\ 1 & \Omega & 1 & \cdots & 0 \\ \vdots & & & & \\ \cdots & 0 & 0 & 1 & \tilde{\Omega} \end{pmatrix}, \tag{52}$$

where we used the abbreviation

$$\tilde{\Omega} := \Omega + i\frac{\delta}{-t}, \tag{53}$$

where $\delta$ is again the external coupling constant. The Green's function can still be found by the same procedure as in the previous section. We here give the final solution:

$$g_{j,m} = \kappa_{j-2}g_{1,m} + \kappa_{j-m-2}, \tag{54}$$

$$g_{1,1} = -\frac{\hat{\kappa}_{n-2}}{\bar{\bar{\kappa}}_{n-1}}, \tag{55}$$

$$g_{1,m} = -\frac{\hat{\kappa}_{n-2}}{\bar{\bar{\kappa}}_{n-1}}\bar{\kappa}_{m-2} + \kappa_{m-3}, \qquad 2 \leq m \leq n-1, \tag{56}$$

$$g_{1,n} = -\frac{1}{\tilde{\Omega}}\left(-\frac{\hat{\kappa}_{n-2}}{\bar{\bar{\kappa}}_{n-1}}\bar{\kappa}_{n-3} + \kappa_{n-4}\right), \tag{57}$$

where the following shortcuts are used:

$$\bar{\kappa}_j := \kappa_j - i\delta\kappa_{j-1} \tag{58}$$

$$\hat{\kappa}_j := \frac{1}{\tilde{\Omega}}\kappa_{j-2} - \frac{i}{\delta}\kappa_j \tag{59}$$

$$\bar{\bar{\kappa}}_j := \frac{1}{\tilde{\Omega}}\bar{\kappa}_{j-2} - \frac{i}{\delta}\bar{\kappa}_j. \tag{60}$$

### C.  Transmission coefficient for the model chain

From the Landauer Theory the transmission coefficient T is finally obtained in terms of the Green's function:

$$T = 4\delta^2 \mid g_{1,n} \mid^2. \tag{61}$$

From Eq. (57) and remembering that g has been chosen so as to contain t in the wake of Eq. (17), $\mid g_{1,n} \mid^2$ can be evaluated to be:

$$\mid g_{1,n} \mid^2 = t^2\frac{[(\kappa_{n-3}\Theta + \kappa_{n-4} + \delta\Gamma\kappa_{n-4})\Omega + \delta(\Gamma\kappa_{n-3} - \delta\Theta)\kappa_{n-4}]^2}{(\Omega^2 + \delta^2)^2}, \tag{62}$$

$$\Theta = -\frac{\delta^2(\kappa_{n-4} + \kappa_{n-2})(\kappa_{n-1} + \kappa_{n-3} - \Omega\kappa_{n-2}) - \Omega\kappa_{n-2}(\delta^2\kappa_{n-4} - \delta\kappa_{n-2} + \Omega\kappa_{n-1})}{\delta^2(\kappa_{n-1} + \kappa_{n-3} - \Omega\kappa_{n-2})^2 + (\delta^2\kappa_{n-4} - \delta\kappa_{n-2} + \Omega\kappa_{n-1})^2},$$

$$\Gamma = \frac{\Omega\delta\kappa_{n-2}(\kappa_{n-1} + \kappa_{n-3} - \Omega\kappa_{n-2}) - \delta(\kappa_{n-4} + \kappa_{n-2})(\delta^2\kappa_{n-4} - \delta\kappa_{n-2} + \Omega\kappa_{n-1})}{\delta^2(\kappa_{n-1} + \kappa_{n-3} - \Omega\kappa_{n-2})^2 + (\delta^2\kappa_{n-4} - \delta\kappa_{n-2} + \Omega\kappa_{n-1})^2}.$$

To give a specific example, we have chosen the parameters $t = 3$ eV, $\delta = 3$ eV and $\epsilon_d = 9$ eV. The number $\eta = n - 2$ denotes the number of units inserted into the model chain except the first and the last element, which serve as coupling elements to the electrodes. The development of T with respect to the number of chain units is shown in Fig. 7. Significant variations with changing $\eta$ are observed. In particular no monotonic decline with the chain length is found. Moreover there is an obvious qualitative agreement between the model case and the curve calculated for the dithiol polyenes with orbital *space 1*, as can be seen from Fig. 6. In both cases there is a monotonic decrease of T with the number of chain units $\eta$ from 1 to 3, followed by an increase for the chain with 4 units.

The fact that this major finding of the *ab initio* calculations can be found in the model treatment as well can be seen as a strong reconfirmation of the former.

## V.  CONCLUSIONS

We have performed a systematic study of the charge transmission behavior of molecules in junctions as a function of the molecule length. In particular the transmission coefficient was computed for $S_2(CH)_{2n}$ for $n \in \{1, 2, 3, 4, 5\}$. The calculations rest on the Landauer theory and the wide band approximation. A fully *ab initio* wave function based procedure was mounted to obtain the Green's function entering the Landauer formula. The local incremental scheme was used with localized occupied, but localized or canonical virtual HF orbitals as a starting point and the results were compared. We found that employing the entire canonical virtual space for excitations leads to a rapid convergence of the incremental contributions to the transmission coefficient, which can be approximated to a large degree by the one–region increments alone. As a result of the series of calculations we find that while T varies with the length $\eta$ of the chain, the dependence is not monotonous. In the second part an analytic formula was developed for the case of a model chain in the tight binding approximation. It turns out that the model analysis can reproduce the qualitative behavior of T as a function of the chain length to a significant extent. We conclude that such a model system might be developed further so as to give qualitative estimates for more complex systems in the future. At the same time this analytic study reconfirms our quantum chemical results for the dithiol polyenes.

## VI.  ACKNOWLEDGMENTS

The authors appreciate support of the German Research Foundation DFG in the frame of the program AL 625/2–1.

---

# Tables

Tab. I

TABLE I: *Correlation correction γ to the HOMO-LUMO gap of the dithiol polyene chains. The obtained values using space 1 and space 2 are given in column three and four respectively. The employed increments are indicated in the second column. The computed dithiol polyenes are designated by the number η of their $H-C{=}C-H$–fragments according to Fig. 1.*

| # $\eta$ of $H-C{=}C-H$–fragments | Increments | γ using *space 1* | γ using *space 2* |
|---|---|---|---|
| 1 | $S$ | 2.900 | 1.215 |
|   | $S + nD$ | 3.039 | 1.815 |
|   | $M$ | 3.032 | 1.839 |
| 2 | $S$ | 2.373 | 0.757 |
|   | $S + nD$ | 2.686 | 1.779 |
|   | $S + nD + nnD$ | 2.810 | 1.858 |
|   | $S + D$ | 2.827 | 1.866 |
|   | $S + D + nT$ | 2.827 | 2.022 |
|   | $S + D + T$ | 2.827 | 2.041 |
|   | $M$ | 2.826 | 2.045 |
| 3 | $S$ | 2.206 | 0.602 |
|   | $S + nD$ | 2.447 | 1.575 |
|   | $S + nD + nnD$ | 2.583 | 1.699 |
|   | $S + nD + nnD + nnnD$ | 2.618 | 1.721 |
|   | $S + D$ | 2.622 | 1.724 |
|   | $S + D + nT$ | 2.626 | 1.902 |
|   | $S + D + T$ | 2.634 | 1.945 |
|   | $M$ | 2.639 | 1.972 |
| 4 | $S$ | 2.104 | 0.519 |
|   | $S + nD$ | 2.298 | 1.449 |
|   | $S + nD + nnD$ | 2.440 | 1.600 |
|   | $S + nD + nnD + nnnD$ | 2.481 | 1.637 |
|   | $S + nD + nnD + nnnD + nnnnD$ | 2.491 | 1.644 |
|   | $S + D$ | 2.492 | 1.645 |
|   | $M$ | 2.521 | 1.923 |
| 5 | $S$ | 2.021 | 0.405 |
|   | $S + nD$ | 2.179 | 1.317 |
|   | $S + nD + nnD$ | 2.334 | 1.484 |
|   | $S + nD + nnD + nnnD$ | 2.380 | 1.533 |
|   | $M$ | 2.445 | 1.878 |

# Figure Captions

Fig. 1 *Illustration of a molecular bridge. The molecule is coupled to the gold electrodes via its sulfur atoms. The Hydrogens have been suppressed in the figure.*

Fig. 2 *Sketch of the incremental scheme, exemplifying a possible partitioning of trans–1,4–dithiolbutadiene in four regions. The upper picture denotes a one–region increment ($S$) in which region I is emphasized by a box, the second panel shows a next–neighbor–two–region increment ($nD$) comprising of regions II and III, and the lower picture represents an increment in which region II and region IV are contained to form a next–next–neighbor–two–region increment ($nnD$). The labeling of other increments is analogous. If the entire molecule is used as one increment, this is donated by $M$.*

Fig. 3 *The transmission coefficient T with varying values of the coupling constant $\delta$ for molecule 4 is depicted by the solid line. The entire molecule M is used as one increment. The dashed line shows the overestimation of T in percent if only all single increments S are included in the calculation. The dotted line depicts this overshooting of T if all S and double increments D are included in the calculation.*

Fig. 4 *Correlation corrected DOS of molecule 4 for a coupling constant $\delta$ of 3 eV.*

Fig. 5 *The transmission coefficient with varying values of the coupling constant $\delta$ from molecule 1 to molecule 5 .*

Fig. 6 *Results of the transmission coefficients from molecule 1 ($\eta = 1$) to molecule 5 ($\eta = 5$) using space 1 and space 2, respectively. The coupling constant $\delta$ is set to 3 eV.*

Fig. 7 *Transmission coefficient T in dependence of the number $\eta$ of inserted units in the model chain.*

# Figures

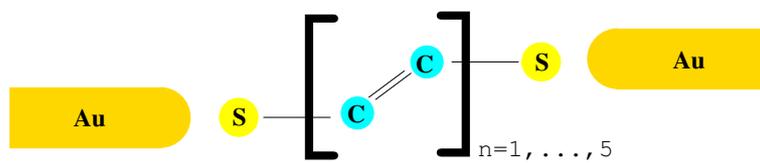

FIG. 1:

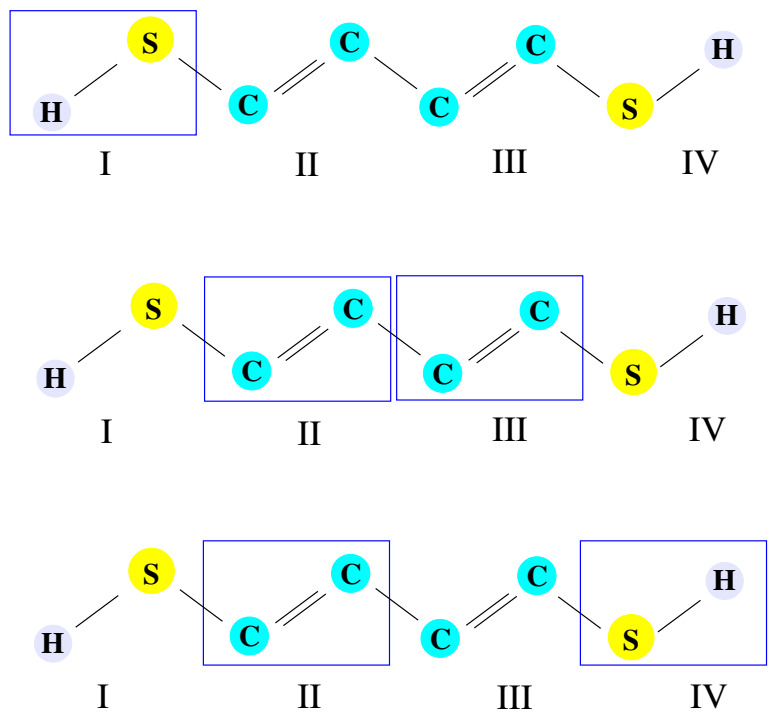

FIG. 2:

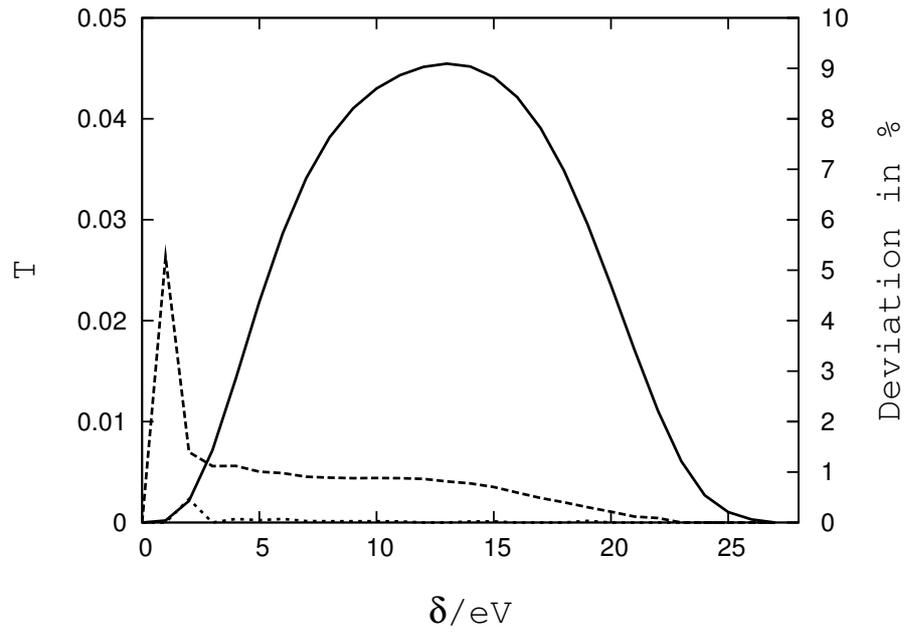

FIG. 3:

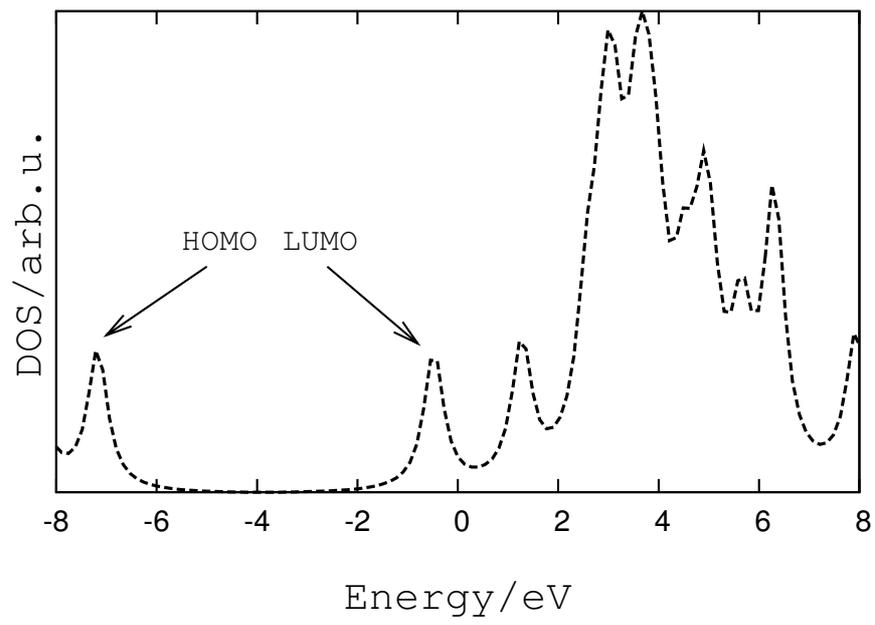

FIG. 4:

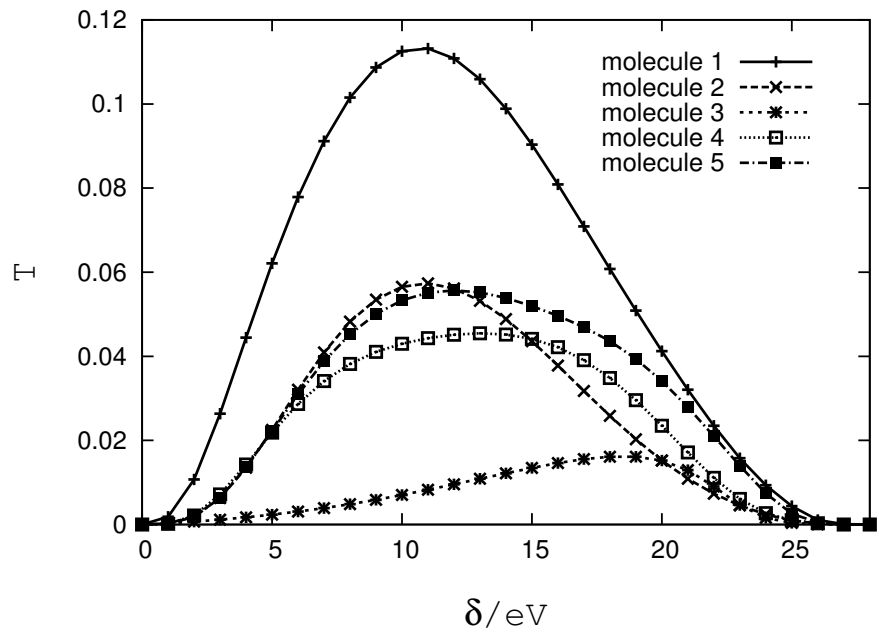

FIG. 5:

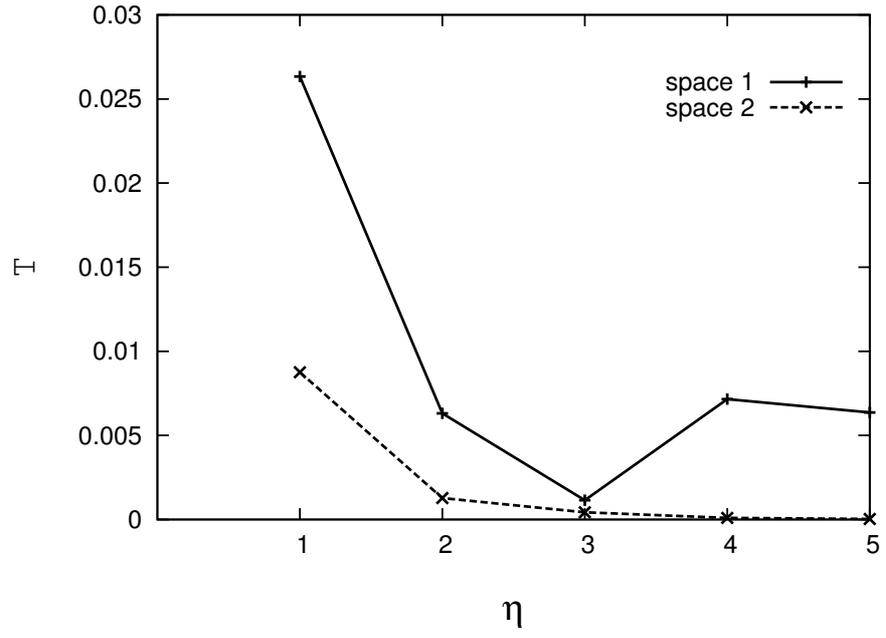

FIG. 6:

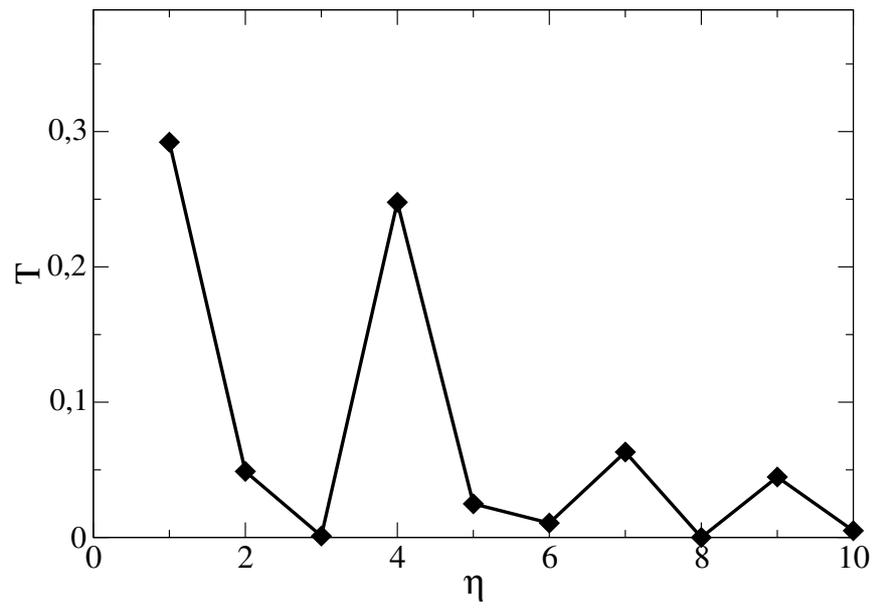

FIG. 7: